\documentclass[manuscript)]{aastex}

\usepackage{lscape,graphicx,epsfig,natbib,color}
\usepackage{epstopdf}

\slugcomment{Submitted to ApJ}

\shorttitle{Chromospheric evaporation in a circular-ribbon flare}
\shortauthors{Zhang et al.}

\begin{document}
\title{Explosive Chromospheric Evaporation in a Circular-ribbon Flare}

\author{Q. M. Zhang\altaffilmark{1,2}, D. Li\altaffilmark{1}, Z. J. Ning\altaffilmark{1}, Y. N. Su\altaffilmark{1}, H. S. Ji\altaffilmark{1}, and Y. Guo\altaffilmark{3,4}}

\affil{$^1$ Key Laboratory for Dark Matter and Space Science,
Purple Mountain Observatory, CAS, Nanjing 210008, China\email{zhangqm@pmo.ac.cn}}
\affil{$^2$ Key Laboratory of Solar Activity, National Astronomical Observatories, CAS, Beijing 100012}
\affil{$^3$ Centre for mathematical Plasma-Astrophysics, Department of Mathematics, KU Leuven, B-3001 Leuven, Belgium}
\affil{$^4$ School of Astronomy and Space Science, Nanjing University, Nanjing 210023, China}

\begin{abstract}
In this paper, we report our multiwavelength observations of the C4.2 circular-ribbon flare in active region (AR) 12434 on 2015 October 16. 
The short-lived flare was associated with positive magnetic polarities and a negative polarity inside, as revealed by the photospheric 
line-of-sight magnetograms. Such magnetic pattern is strongly indicative of a magnetic null point and spine-fan configuration in the corona.
The flare was triggered by the eruption of a mini-filament residing in the AR, which produced the inner flare ribbon (IFR) and the southern part of a closed circular flare ribbon (CFR). 
When the eruptive filament reached the null point, it triggered null point magnetic reconnection with the ambient open field and generated the bright CFR and a blowout jet.
Raster observations of the \textit{Interface Region Imaging Spectrograph} (\textit{IRIS}) show plasma upflow at speed of 35$-$120 km s$^{-1}$ in the Fe {\sc xxi} 1354.09 {\AA} line ($\log T\approx7.05$) 
and downflow at speed of 10$-$60 km s$^{-1}$ in the Si {\sc iv} 1393.77 {\AA} line ($\log T\approx4.8$) at certain locations of the CFR and IFR during the impulsive phase of flare, indicating explosive chromospheric evaporation. 
Coincidence of the single HXR source at 12$-$25 keV with the IFR and calculation based on the thick-target model suggest that the explosive evaporation was most probably driven by nonthermal electrons.
\end{abstract}

Online-only material: animations, color figures

\section{Introduction}
Solar flares are impulsive increases of emissions in various wavelengths from radio to hard X-ray (HXR) in the solar atmosphere \citep{benz08,fle11}. 
Within tens of minutes to several hours, the free magnetic energies (10$^{29}$$-$10$^{32}$ ergs) accumulated before the flares are released and converted 
into the kinetic and thermal energies via magnetic reconnection \citep{pri00,su13}. In the thick-target model, the accelerated nonthermal electrons 
(20$-$100 keV) are precipitated downward in the much denser chromosphere, leading to impulsive heating of the local plasma up to 
$\sim$10 MK and rapid increase in HXR emissions via Coulomb collisions \citep{bro71,cheng10}. When the heating timescale is much shorter 
than the radiative cooling timescale, the overpressure of the chromosphere pumps hot plasma into the newly reconnected coronal loops that emit strong 
emissions in extreme ultraviolet (EUV) and soft X-ray (SXR), a process called chromospheric evaporation \citep{fis85a,fis85b,fis85c,mar89,ems92,abb99,all05,all15}. 
So far, it has been extensively investigated using both HXR imaging \citep{liu06,ning09} and spectroscopic observations in H$\alpha$, EUV, and SXR 
wavelengths \citep{act82,cza99,bro04,you13,pol15,pol16}. \citet{liu06} studied the spatial evolution of the HXR emissions during the impulsive phase of 
an M1.7 flare. They found that the HXR emission centroids move from the footpoint toward the loop top at speed of hundreds of km s$^{-1}$, 
which is indicative of continuous chromospheric evaporation as a result of the deposition of electron energies.

Chromospheric evaporations are divided into two types according to the level of energy flux \citep{fis85b}. Explosive evaporation occurs when the input energy 
flux exceeds the critical value ($\sim$10$^{10}$ erg cm$^{-2}$ s$^{-1}$). Plasma upflows at speed of hundreds of km s$^{-1}$ are observed in the emission lines 
formed in the coronal temperatures, while downflows at speed of tens of km s$^{-1}$ are observed in the emission lines formed in the transition region and upper 
chromosphere \citep{bro04,mil06b}. The total momentum of upflowing plasma is approximately equal to that of the downflowing plasma or chromospheric 
condensation \citep{fis87,can90}. Otherwise, gentle evaporations take place accompanied by upflows at speed of tens of km s$^{-1}$ in all emission 
lines \citep{mil06a,sad15}. Recently, \citet{reep15} investigated the importance of electron energy on the two types of evaporation. They found that the
threshold between explosive and gentle evaporation is not fixed at a given beam energy flux. Instead, it depends strongly on the electron energy and duration of heating.
Occasionally, conversions from impulsive type to gentle type or from gentle type to explosive type are observed in different phases of flares \citep{bro09,li15b}. 

Apart from the nonthermal electrons, thermal conduction also plays a role in driving chromospheric evaporations \citep{zar88,bat09,zqm13}, which has been explored in 
magnetohydrodynamic (MHD) numerical simulations \citep[e.g.,][]{yoko98,bra14,long14}. 
\citet{reep16} discovered that Alfv\'{e}nic waves, propagating from the corona to the chromosphere, can also heat the upper chromosphere and produce explosive evaporation.
To date, the dominant driving mechanism is still controversial \citep{wue94,raf09}. The successful launch of the \textit{Interface Region Imaging Spectrograph} 
\citep[\textit{IRIS};][]{dep14} telescope opened a new era for the study of flare dynamics. Evidences of electron-driven chromospheric evaporations have 
been reported using state-of-the-art observations of \textit{IRIS} \citep[e.g.,][]{bat15,gra15,li15a,li15b,tian15}.

In the context of standard flare and coronal mass ejection (CME) model \citep[e.g.,][]{shi95,lin04}, there are two parallel flare ribbons where nonthermal electrons collide and heat 
the chromosphere, which are observed in the Ca {\sc ii} H, H$\alpha$, UV, and EUV wavelengths. Apart from the two ribbons, a particular type of flare ribbons, i.e. circular ribbons, 
exist \citep{mas09,reid12,wang12,jia13,liu13,sun13,yang15,zqm15}. They are always associated with the spine-fan configuration in the presence of magnetic null point, 
which is a singular point where the magnetic field vanishes ($\mathbf{B=0}$) \citep{lau90}. The magnetic field $\mathbf{B}$ near the null point can 
be expressed as the linear term $\mathbf{B=M\cdot r}$, where $\mathbf{M}$ is a Jacobian matrix with elements $M_{ij}=\partial B_{i}/\partial x_{j}$ and $\mathbf{r}$ is the 
position vector $(x,y,z)^T$ centered at the null point \citep{par96}. The divergence-free condition ($\mathbf{\nabla \cdot B=0}$) requires that the sum 
of the three eigenvalues equal to zero \citep{zqm12}. The two eigenvectors corresponding to the two eigenvalues of the same sign determine the fan surface, which divides the 
space into two regions having a distinct connectivity. The third eigenvector corresponding to the third eigenvalue of the opposite sign determines the direction of spines passing 
through the null point. Magnetic reconnection and particle 
acceleration in null point reconnection regions have been explored in analytical study \citep{pri96,lit04} and three-dimensional (3D) numerical simulations \citep{ros10,bau13a,bau13b}.
The circular ribbons are believed to be intersections of the fan surfaces and the chromosphere. The central or inner ribbons within the circular ribbons are thought to be intersections 
of the inner spines and the chromosphere \citep{wang12,reid12}. Sometimes, there are multiple flare ribbons owing to the extraordinarily complex magnetic topology of the 
active regions \citep[ARs;][]{jos15,liu15}. 

So far, chromospheric evaporations in circular-ribbon flares have rarely been observed and investigated, especially by \textit{IRIS}.
Although significant improvements in understanding the chromospheric evaporations have been achieved, there are still open questions need to be addressed: 
How is the circular-ribbon flares generated? Are there explosive or gentle chromospheric evaporations in circular-ribbon flares? What is the cause of evaporation?
In this paper, we report the multiwavelength imaging and spectral observations of the \textit{GOES} C4.2 circular-ribbon flare in NOAA AR 12434 (S10E37), 
which is one of the homologous flares on 2015 October 16. Data analysis and results of the filament eruption and flare are shown in Section~\ref{s-flare}. 
Data analysis and results of the explosive chromospheric evaporations in the circular flare ribbon (CFR) and inner flare ribbon (IFR) are presented 
in Section~\ref{s-evap}. Discussion and summary are arranged in Section~\ref{s-disc} and Section~\ref{s-sum}.

\section{Filament eruption and circular-ribbon flare} \label{s-flare}

\subsection{Observation and data analysis} \label{s-data1}
The flare was observed by the ground-based telescope of the Global Oscillation Network Group (GONG) in H$\alpha$ line center and by the Atmospheric Imaging 
Assembly \citep[AIA;][]{lem12} aboard the \textit{Solar Dynamic Observatory} (\textit{SDO}) in 1600 {\AA} and EUV wavelengths (94, 131, 171, 193, 211, 304, 335 {\AA}).
The photospheric line-of-sight (LOS) magnetograms were observed by the Helioseismic and Magnetic Imager \citep[HMI;][]{sch12} aboard \textit{SDO}. 
The level\_1 data from AIA and HMI were calibrated using the SSW programs \textit{aia\_prep.pro} and \textit{hmi\_prep.pro}, respectively.  To locate where the nonthermal 
particles precipitate, we made HXR images using the CLEAN method with integration time of 120 s at different energy bands of the \textit{Reuven Ramaty High-Energy 
Solar Spectroscopic Imager} \citep[\textit{RHESSI};][]{lin02}. The images observed 
in H$\alpha$, UV, EUV, and HXR wavelengths were coaligned with accuracy of $\sim$0$\farcs$6. The observing parameters of the instruments are summarized in Table~\ref{tbl-1}.

\subsection{Results} \label{s-result1}
\begin{figure}
\epsscale{.80}
\plotone{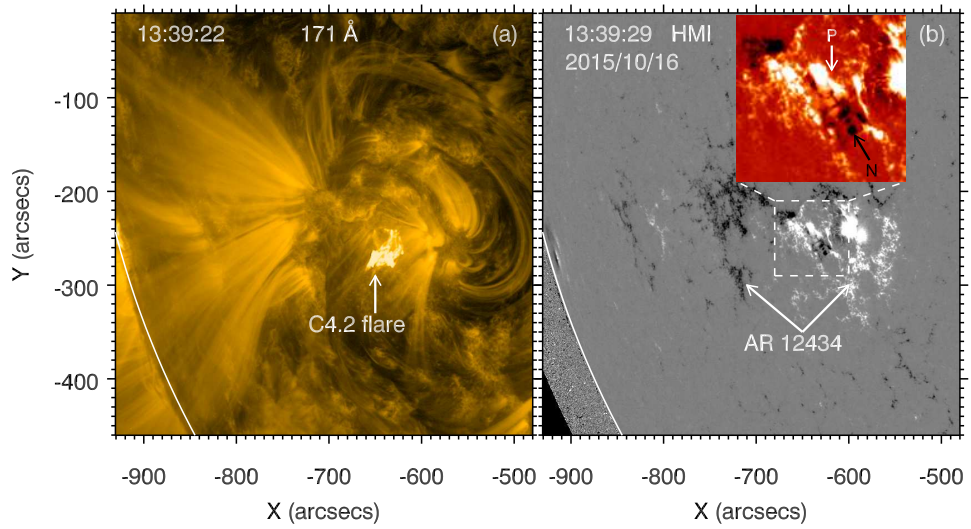}
\caption{(a) EUV 171 {\AA} image observed by AIA at 13:39:22 UT. The white arrow points to the cicular-ribbon flare.
(b) Photospheric LOS magnetogram observed by HMI at 13:39:29 UT. The white arrows point to the main negative and positive polarities of the AR. 
The inset colored image shows the close-up of the flare region within the white dashed box.
\label{fig1}}
\end{figure}

Figure~\ref{fig1} shows the 171 {\AA} image observed by AIA and the LOS magnetogram observed by HMI at $\sim$13:39 UT. The flare occurred in the AR 12434, 
which is characterized by large-scale coronal arcade. In panel (b), the inset colored image shows the close-up of the flare region, which is characterized by a central 
negative polarity (N) surrounded by the positive polarities (P).

Figure~\ref{fig2} shows the temporal evolution of the flare in 304 {\AA}. Before the flare, a very small filament, which is indicated by the arrows, resided in the AR (see panels (a)-(b)). 
As time goes on, the dark mini-filament activated and generated elongated, jet-like brightening at $\sim$13:35. After 13:37:30 UT, the filament erupted impulsively and generated the 
C4.2 flare, which features a CFR (see panels (e)-(f) and online movie \textit{anim304.mpg}). Meanwhile, the cool filament was heated significantly. 
In panel (f), the brightest region in the southwest of CFR coincides with the single HXR source at 12$-$25 keV. The rising and expanding filament became a curtain-like, blowout jet after 13:39:30 UT.
\citet{moo10} classified the coronal jets into the standard type and blowout type. The standard type has simpler morphology and can be explained by the magnetic emerging-flux model \citep{shi92}. 
The blowout type, however, results from small-scale filament eruptions accompanied by rotating and/or transverse drifting motions \citep[e.g.,][]{par09,moo13,zqm14,kum16}.
In our work, the jet not only propagated longitudinally, but also underwent transverse drift from west to east. Interestingly, the brightening at the base of jet propagated in the counterclockwise 
direction along the CFR. In panel (g), the contours of the positive and negative LOS magnetic fields 
are superposed with magenta and green lines, respectively. As mentioned above in Figure~\ref{fig1}(b), the negative polarity (N) is surrounded by positive polarities (P), which strongly implies 
the existence of magnetic null point and spine-fan configuration in the upper atmosphere \citep{zqm12,zqm15}. The CFR is approximately cospatial with the positive polarities, while the IFR is 
approximately cospatial with the negative polarity. In panels (h) and (i), the short and bright IFR within the CFR is cospatial with the HXR source.

The evolution of flare observed in H$\alpha$ is displayed in the top panels of Figure~\ref{fig3} with lower resolution. The pre-existing dark filament, which was $\sim$15$\arcsec$ away 
from the sunspot, kept stable until $\sim$13:35 UT. It erupted and generated the blowout jet and cicular-ribbon flare (see the online movie \textit{animha.mpg}). The filament eruption and flare 
were also evident in the other wavelengths of AIA with higher formation temperatures. The rest panels of Figure~\ref{fig3} demonstrate selected images observed by AIA. 
The 335 {\AA} ($\log T\approx6.4$) image and 94 {\AA} ($\log T\approx6.8$) image are characterized by the hot post flare loops (PFLs), which connect the IFR and southern part of CFR.
The eastern part of CFR is less evident in 94 {\AA} than in the other wavelengths, implying that the temperatures of the eastern part of CFR were lower than the western part.

In order to investigate the evolution of the blowout jet, we selected two slices in Figure~\ref{fig3}(h): S1 along the jet axis and S2 across the axis. The time-slice 
diagrams of S1 and S2 in 171 {\AA} are displayed in the left and right panels of Figure~\ref{fig4}, respectively. The jet started at $\sim$13:39 UT and propagated 
outwards along the axis at a speed of $\sim$308 km s$^{-1}$. Meanwhile, the jet underwent transverse drifting motion from west to east at a speed of 
$\sim$87 km s$^{-1}$, which is much higher than the previously reported values \citep{moo13}.

In Figure~\ref{fig5}, the upper panel shows the SXR light curves during 13:20$-$14:00 UT in 0.5$-$4 {\AA} and 1$-$8 {\AA}. The short-lived C4.2 flare had lifetime of 
$\sim$15 minutes. It started at $\sim$13:36:30 UT, peaked at $\sim$13:42:31 UT, and ended at $\sim$13:51 UT. The HXR light curves at various energy bands
(3$-$6, 6$-$12, 12$-$25, 25$-$50, 50$-$100 keV) are plotted with colored lines in the lower panel. The peak times at HXR energy bands preceded the SXR peak time by 1$-$2 minutes, 
implying the Neupert effect \citep{ning10}.

\section{Chromospheric evaporation in flare ribbons} \label{s-evap}

\subsection{Observation and data analysis} \label{s-data2}
Fortunately, the flare was captured by \textit{IRIS} Slit-Jaw Imager (SJI) in 1400 {\AA} and raster observation 
in the ``sparse synoptic raster'' mode. Each raster had 36 steps from east to west and covered an area of 35$\farcs$5$\times$181$\farcs$5. The step cadence and 
exposure time were $\sim$9.4 s and 7.1 s. Each step had a spatial size of $\sim$0$\farcs$166 and a spectral scale of $\sim$25.6 m{\AA} pixel$^{-1}$ in the far 
ultraviolet bands, which equals to $\sim$5.7 km s$^{-1}$ pixel$^{-1}$. The sixth raster data of Fe {\sc xxi} and Si {\sc iv} lines during 13:37:29$-$13:43:00 UT were 
preprocessed using the standard Solar Software (SSW) programs \textit{iris\_orbitvar\_corr\_l2.pro} and \textit{iris\_prep\_despike.pro}. The Fe {\sc xxi} line 
($\log T\approx7.05$) is blended with cold and narrow chromospheric lines, which should be identified and removed using the multi-Gaussian fitting method \citep{li15a,li16}.
The line centers and widths of these blended lines are fixed or constrained, while their intensities are tied to specific species in adjacent spectral window. 
The raster observations in this study lacked the spectral window ``1343'', which includes the tied line at H$_2$ 1342.77 {\AA}. The two blended lines at 1353.32 {\AA} 
and 1353.39 {\AA} were too weak to contribute to the Fe {\sc xxi}, so that they were not considered in this fitting. The Si {\sc iv} line ($\log T\approx4.8$) is an isolated line, 
which can be well fitted by the single-Gaussian function.

\subsection{Results} \label{s-result2}
Figure~\ref{fig6} shows the 1400 {\AA} images observed by \textit{IRIS}/SJI with extremely high resolution during 13:31:23$-$13:46:28 UT 
(see online movie \textit{anim1400.mpg}). The evolution of the flare is quite similar to that in 304 {\AA} in Figure~\ref{fig2}, featuring bright CFR and IFR with ultrafine structures.
The intensity of CFR did not increase simultaneously but in the counterclockwise direction. Like in EUV and H$\alpha$ wavelengths, the single HXR source is exactly located at the IFR (see panel (g)). 
The two vertical dashed lines in panel (e) denote the starting and ending positions of the 36-step raster observation, exactly covering the flare and jet during the impulsive phase.
Since the slit and CFR intersects in two places during the scan, we call the northern and southern intersections NCFR and SCFR, respectively. A few selected 
points at the NCFR, SCFR, and IFR are displayed as green, blue, and magenta pluses in panel (f).

The \textit{IRIS} spectra windows of Fe {\sc xxi} and Si {\sc iv} at three times are shown in the left and right panels of Figure~\ref{fig7}, respectively. The spectra
profiles and results of multi-Gaussian fitting of three points representative of NCFR, IFR, and SCFR are displayed in the left panels. The fitting results of Fe {\sc xxi} 
spectra are drawn in turquoise lines. It is clear that the line centers of the three points are blueshifted compared with the rest wavelength of Fe {\sc xxi} at 1354.09 {\AA} 
\citep{li15a,li16}, indicating upflows of super-hot plasma.
The spectra profiles and results of single-Gaussian fitting of the same points are demonstrated in the right panels. The spectra of a nonflaring region is used for 
determining the rest wavelength of Si {\sc iv} (1393.77 {\AA}). It is evident that the line centers of the three points are redshifted compared with the rest wavelength, 
indicating downflows of plasma with temperature of $\sim$0.063 MK.

The calculated Doppler velocities of the NCFR, SCFR, and IFR in Figure~\ref{fig6}(f) are plotted with diamonds, crosses, and boxes with error bars in Figure~\ref{fig8}, 
respectively. The simultaneous upflows at speed of 35$-$120 km s$^{-1}$ in the high-temperature line and downflows at speed of 10$-$60 km s$^{-1}$ 
in the low-temperature line suggest that explosive chromospheric evaporation took place in the CFR and IFR during the impulsive phase of flare (13:39$-$13:43 UT).

\section{Discussion} \label{s-disc}

\subsection{How is the circular-ribbon flare generated?} \label{s-erupt}
\begin{figure}
\epsscale{.70}
\plotone{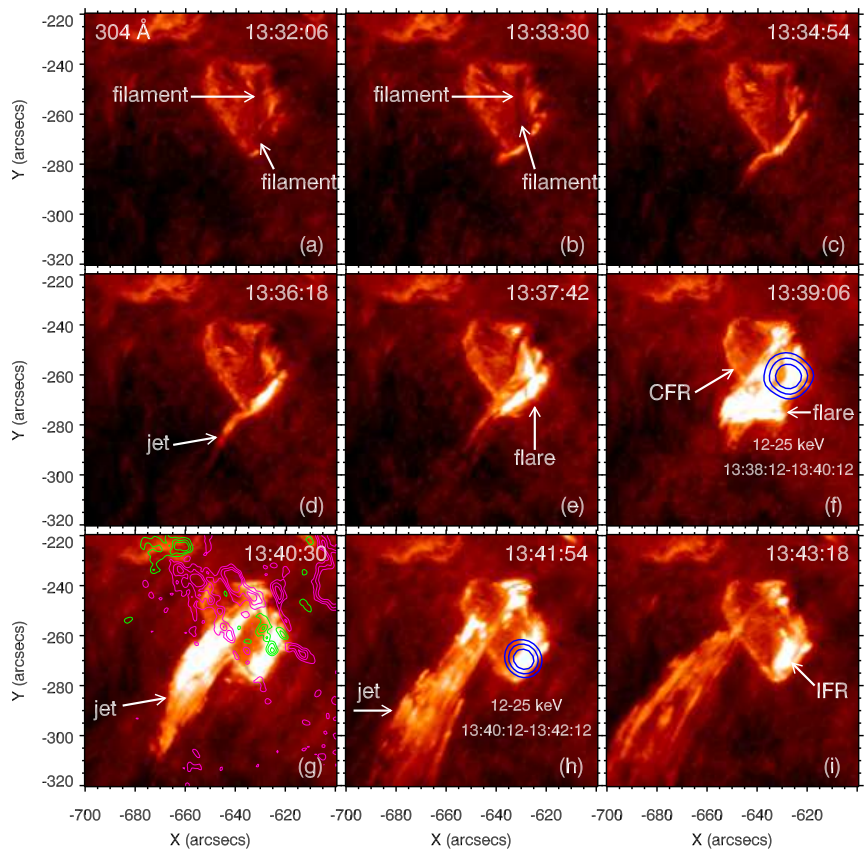}
\caption{Snapshots of the EUV 304 {\AA} images observed by AIA. The white arrows point to the dark filament, jet, CFR, and IFR. 
In panel (g), the contours of the positive and negative LOS magnetic fields are superposed with magenta and green lines, respectively.
In panels (f) and (h), the contours of the HXR images with levels of 70\%, 80\%, and 90\% of the maximum intensity are superposed with blue lines.
\label{fig2}}
(Animations of this figure are available in the online journal.)
\end{figure}

\begin{figure}
\epsscale{.80}
\plotone{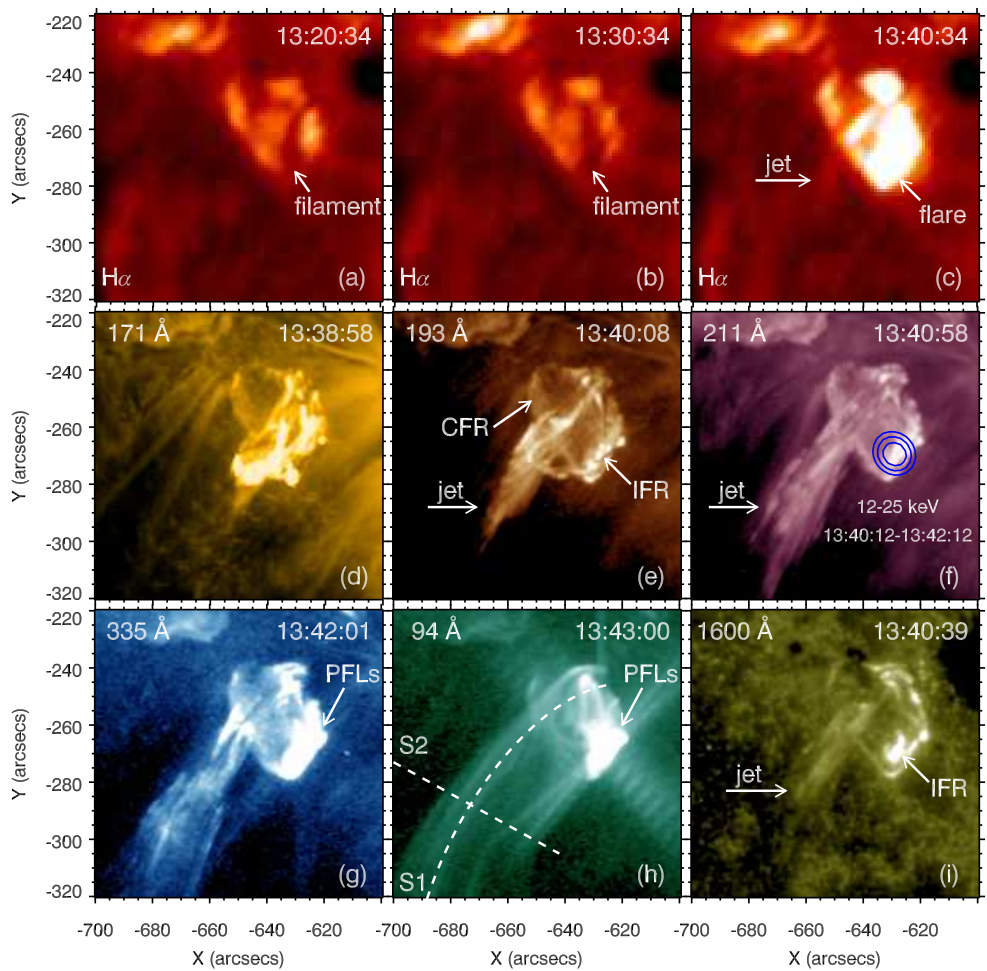}
\caption{(a)-(c) Snapshots of the H$\alpha$ images. The dark filament, which is $\sim$15$\arcsec$ away from the AR 
sunspot, is pointed by the white arrows in panels (a)-(b). In panel (c), the white arrows point to the bright flare and jet. 
(d)-(i) Snapshots of the EUV and UV images observed by AIA. The white arrows point to the jet, CFR, IFR, and post flare loops (PFLs).
In panel (f), the intensity contours of the HXR image are superposed with blue lines.
In panel (h), two slices S1 and S2 are used to investigate the longitudinal and transverse evolutions of the blowout jet.
\label{fig3}}
(Animations of this figure are available in the online journal.)
\end{figure}

\begin{figure}
\epsscale{.80}
\plotone{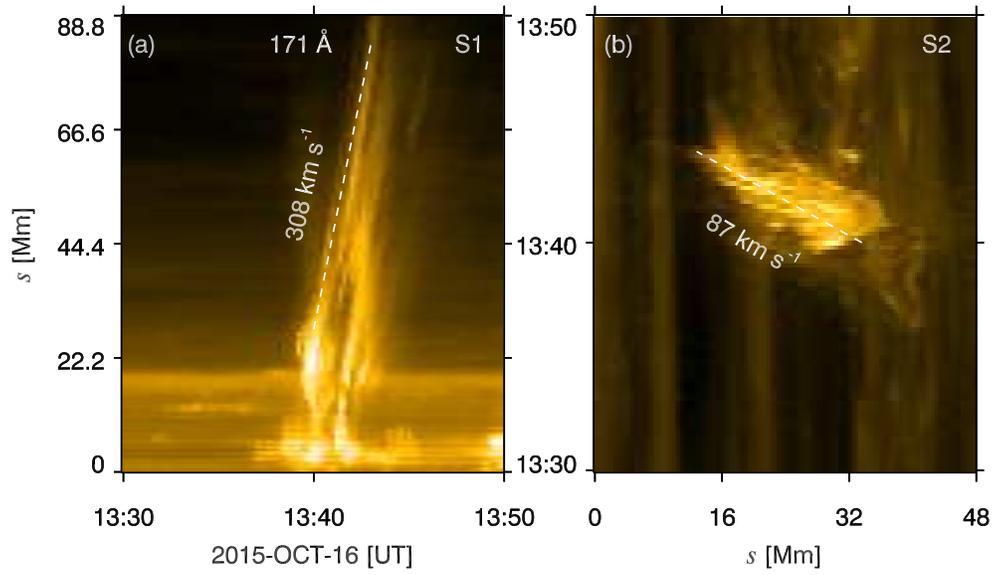}
\caption{Time-slice diagrams of S1 and S2 in 171 {\AA}. $s=0$ stands for the northwest and northeast endpoints for S1 and S2, respectively.
The apparent propagation velocity ($\sim$308 km s$^{-1}$) and transverse drift velocity ($\sim$87 km s$^{-1}$) are also labeled.
\label{fig4}}
\end{figure}

\begin{figure}
\epsscale{.60}
\plotone{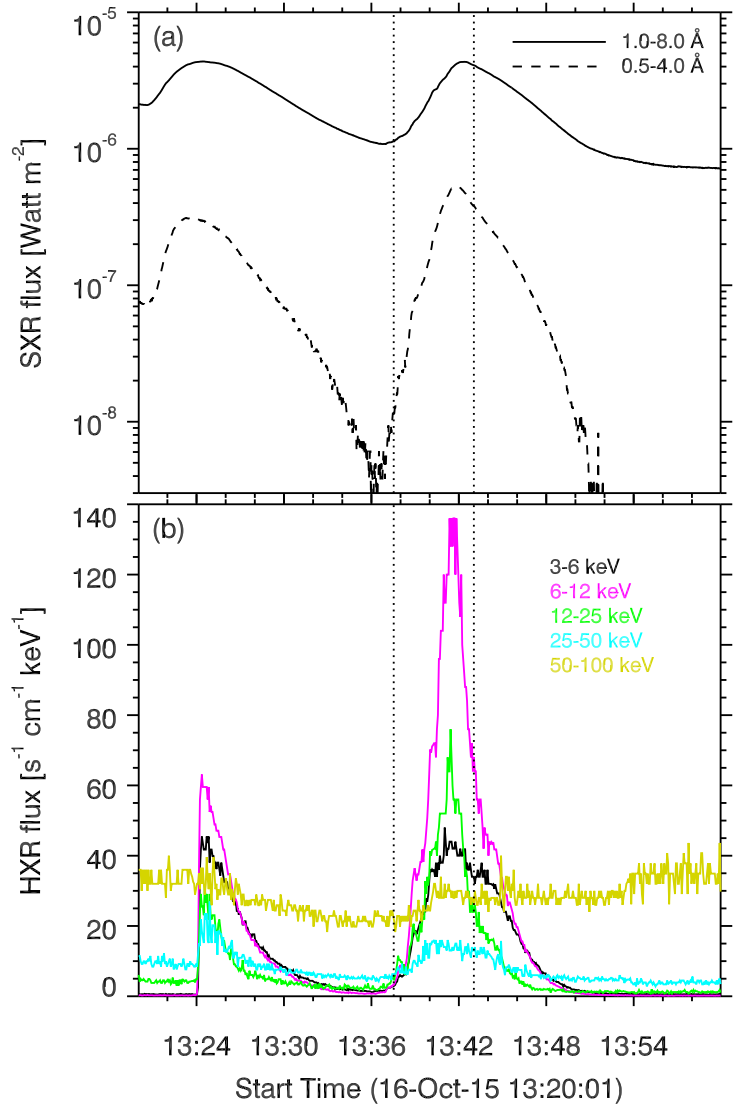}
\caption{(a) SXR light curves during 13:20$-$14:00 UT in 0.5$-$4 {\AA} (\textit{dashed line}) and 1$-$8 {\AA} (\textit{solid line}). 
(b) HXR light curves at various energy bands. 
The two dotted lines denote the starting (13:37:29 UT) and ending (13:43:00 UT) times of the \textit{IRIS} raster observation we used.
\label{fig5}}
\end{figure}

\begin{figure}
\epsscale{.80}
\plotone{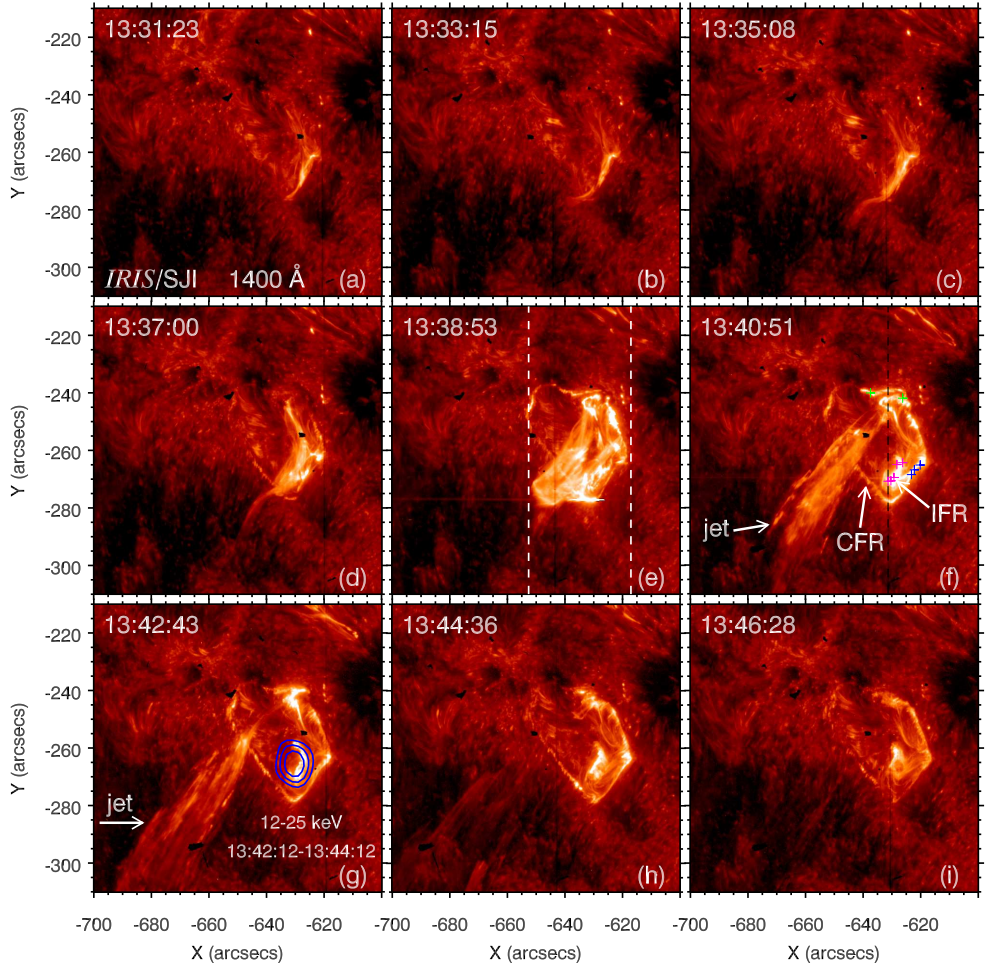}
\caption{Snapshots of the \textit{IRIS}/SJI 1400 {\AA} images. In panel (e), the two vertical dashed lines denote the starting and ending positions 
of the raster observation. In panel (f), the white arrows point to the CFR, IFR, and jet. The vertical dashed line denote the slit position at 13:40:49 UT.
The selected positions of the NCFR, IFR, and SCFR are labeled with green, magenta, and blue pluses. In panel (g), the intensity contours of the HXR image 
are superposed.
\label{fig6}}
(Animations of this figure are available in the online journal.)
\end{figure}

\begin{figure}
\epsscale{.70}
\plotone{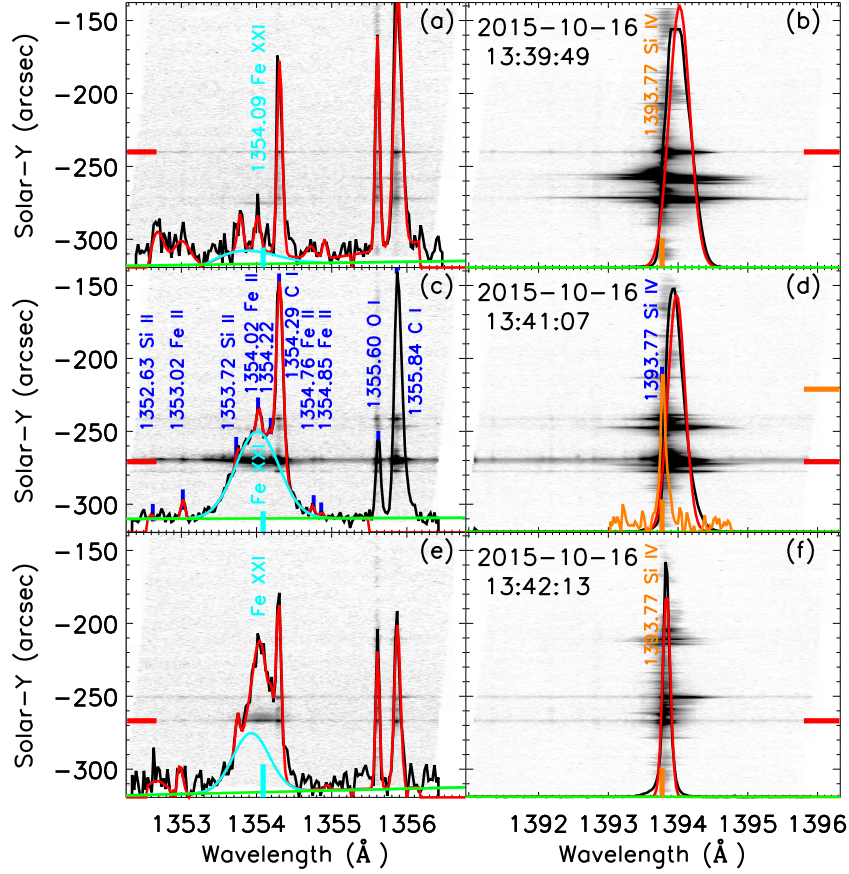}
\caption{\textit{IRIS} spectra windows (\textit{left column} for ``Fe {\sc xxi}'' and \textit{right column} for ``Si {\sc iv}'') at 13:39:49 UT (\textit{top row}), 13:41:07 UT 
(\textit{middle row}), and 13:42:13 UT (\textit{bottom row}). In each panel, the black curve is the spectra at the location marked by the red horizontal line.
In left panels, the red curves represent the results of multi-Gaussian fitting, and the turquoise profiles are Fe {\sc xxi}. The rest wavelength (1354.09 {\AA}) is 
labeled with turquoise vertical ticks. The ten blended lines are labeled with blue vertical ticks in panel (c).
In right panels, the red curves represent the results of single-Gaussian fitting. The orange curve is the spectra for the nonflaring region, which is used for determining 
the rest wavelength of Si {\sc iv}, i.e., 1393.77 {\AA}.
\label{fig7}}
\end{figure}

\begin{figure}
\epsscale{.80}
\plotone{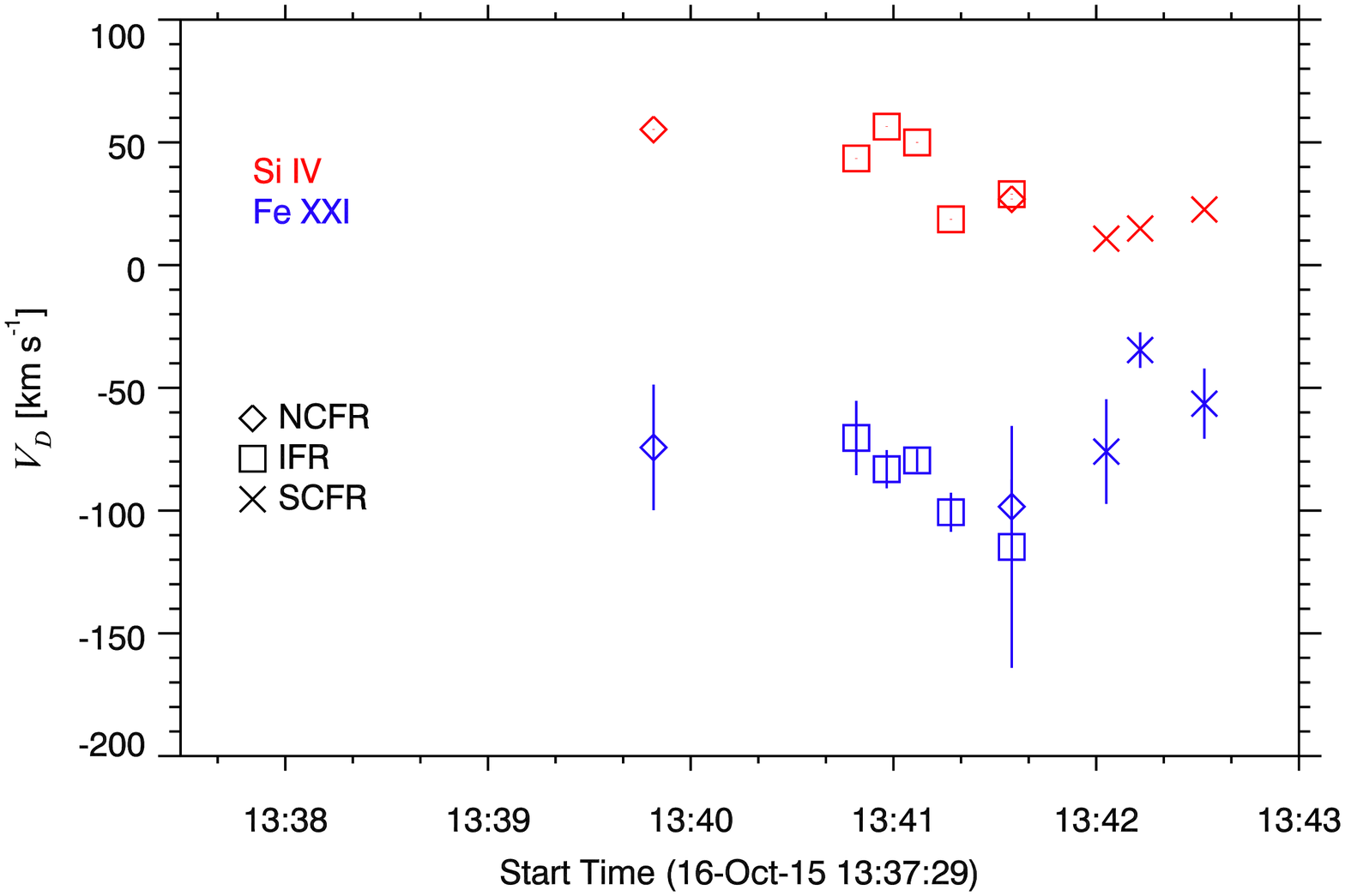}
\caption{Doppler velocities of the downflow derived from the Si {\sc iv} 1393.77 {\AA} line and upflow derived from the Fe {\sc xxi} 1354.09 {\AA} 
line for the NCFR (\textit{diamonds}), IFR (\textit{boxes}), and SCFR (\textit{crosses}). The error bars of the velocities are indicated.
\label{fig8}}
\end{figure}

Owing to the rapid increases of spatial resolutions and observational data of the space-borne telescopes, more and more circular-ribbon flares have been observed 
and reported \citep[e.g.,][]{mas09,reid12,wang12}. \citet{jos15} studied the M7.3 flare as a result of sigmoid eruption in a large-scale fan-spine-type magnetic configuration 
on 2014 April 18. The flare consists of parallel ribbons and a large-scale quasi-circular ribbon. To explain the observational aspects, the authors use a multi-step 
magnetic reconnection: tether-cutting reconnection for the formation and eruption of the sigmoid, standard reconnection in the wake of the erupting sigmoid for the parallel ribbons, 
and null-type reconnection for the quasi-circular ribbon and blowout jet, which is a 3D breakout-type eruption in nature and has been studied in the previous numerical 
simulations of flux rope eruptions and CMEs \citep{lyn08,lyn09}.
In our work, the eruption could be understood as follows. First, the mini-filament became unstable and rose as a result 
of tether-cutting reconnection, magnetic flux emergence, or ideal MHD instability \citep{zqm15}, which is out of the scope of this paper and will be the main topic of the next paper. 
The slow activation was accompanied by small-scale brightening and jet-like motion (see Figure~\ref{fig2}(b)-(d)).
The reconnection in the wake of the erupting filament generated the IFR and SCFR, which can be considered as parallel two ribbons (see Figure~\ref{fig2}(e)). 
When the filament reached the null point, it reconnected with the ambient field and produced the blowout jet and bright CFR (see Figure~\ref{fig2}(f)-(i)). 
The accumulated twist in the filament was transferred to the ambient open field during the magnetic reconnection \citep{par09,more13}. 
This is consistent with the transverse drift of the jet and the sequential brightening of CFR in the counterclockwise direction (see Figure~\ref{fig4}(b) and Figure~\ref{fig6}).

\subsection{What is the cause of chromospheric evaporation?} \label{s-cause}
Chromospheric evaporation is an important process in flare dynamics and has been extensively investigated. Explosive chromospheric evaporations in two-ribbon 
flares have been observed and reported \citep[e.g.,][]{cza99,bat15,gra15,li15b,tian15}. \citet{li15a} explored the relationship between HXR emissions and Doppler velocities 
caused by the explosive chromospheric evaporation in two X1.6 flares on 2014 September 10 and October 22. The correlations between the HXR emissions and 
Doppler shifts of Fe {\sc xxi} and C {\sc i} ($\log T\approx4.0$) suggest that the explosive evaporations in the flares are driven by electrons. 
\citet{bat15} studied the chromospheric evaporations of the X1.0 flare on 2014 March 29. They found that the locations of HXR footpoint sources were coincident 
with the locations of upflow in part of the southern ribbon during the peak of the flare. During the decay phase, the evaporation was probably driven by energy flux via
thermal conduction. They concluded that electron beam may play a role only in driving the chromospheric evaporation during the initial phases of flare.
In our study, explosive evaporation took place not only in the CFR, but also in the IFR. The single HXR source was cospatial with 
the IFR, meaning that the explosive evaporation was most probably driven by nonthermal electrons accelerated by the flare (see Figure~\ref{fig2}, Figure~\ref{fig3}, and Figure~\ref{fig6}). 
It should be emphasized that the integration time of the HXR images is 120 s, which is longer than that in \citet{bat15}. We have tried making HXR images using 60 s integration 
time and found that the time and location of HXR source are the same. The HXR images, however, became more dispersive due to the lower photon count rate and signal-to-noise ratio.

\begin{figure}
\epsscale{.80}
\plotone{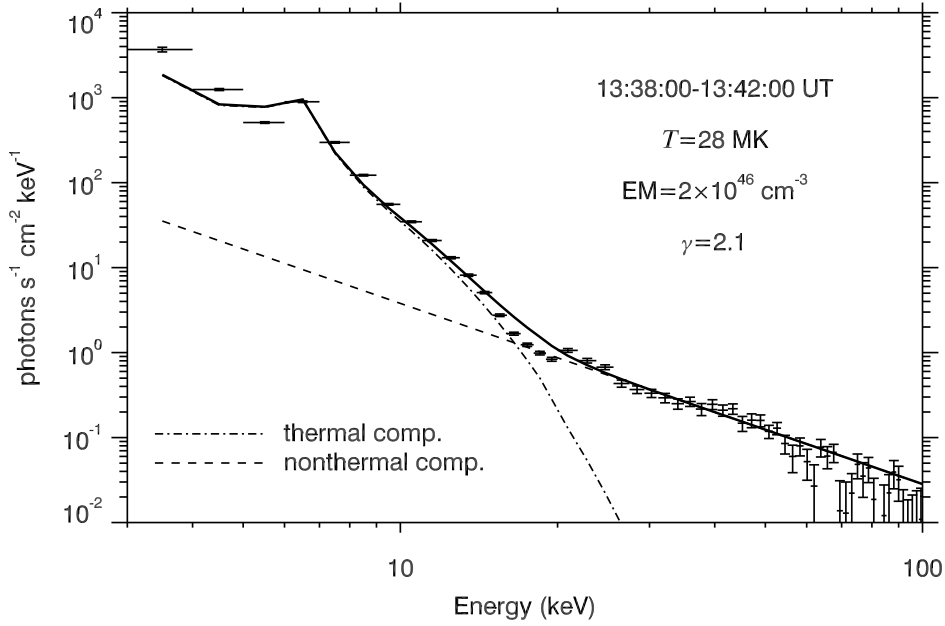}
\caption{Results of \textit{RHESSI} spectral fitting during 13:38$-$13:42 UT on 2015 October 16. The data points with horizontal and vertical error bars represent the
observational data. The spectra for the thermal component and power-law nonthermal component are drawn with dot-dashed line and dashed line, respectively. The sum 
of both components is drawn with thick solid line. The integration time and values of fitted parameters, including the thermal temperature ($T$), emission measure (EM), 
and power-law index ($\gamma$), are displayed.
\label{fig9}}
\end{figure}

In order to justify our conjecture of electron-driven evaporation, we made HXR spectrum during the impulsive phase of flare. The 4-min integration time (13:38$-$13:42 UT) 
is sufficient to get a higher signal-to-noise ratio and smaller error bars. The spectrum and results of two-component fitting are displayed in Figure~\ref{fig9}. The spectra for 
thermal component and power-law nonthermal component are drawn in dot-dashed and dashed lines. The thermal temperature ($T$) and emission measure (EM) 
are $\sim$28 MK and 2$\times10^{46}$ cm$^{-3}$. The power-law slope or spectral index ($\gamma$) of the HXR photons is $\sim$2.1. Therefore, the electron 
spectral index $\delta=\gamma+1=3.1$. The total nonthermal power $P_{tot}$ above a cut-off energy $E_c$ is 1.16$\times10^{24} \gamma^3I_{1}(E_{c}/E_{1})^{-(\gamma-1)}$ erg s$^{-1}$,
where $I_{1}$ denotes the photon count rate \citep{asch04}.
For the C4.2 flare, assuming that $I_{1}=10^3$ photon s$^{-1}$ cm$^{-2}$ and $E_{c}=E_{1}=20$ keV, $P_{tot}$ is estimated to be 1.1$\times10^{28}$ erg s$^{-1}$.
Considering that the area of HXR source is in the range of $2.6\times10^{17}-1.1\times10^{18}$ cm$^2$ (see Figure~\ref{fig2}(h)), the total nonthermal energy flux is 1$\times10^{10}$$-$4$\times10^{10}$ erg s$^{-1}$ cm$^{-2}$, 
which is greater than the threshold for explosive chromospheric evaporation \citep{fis85b}. Therefore, the explosive evaporation in our study is most probably driven by 
nonthermal electrons. \citet{pol15} studied the C6.5 flare on 2014 February 3 
and found that blueshifted ($>$80 km s$^{-1}$) profiles of the Fe {\sc xxi} appear at the very early phase of flare and gradually decrease to 15 km s$^{-1}$ in $\sim$6 
minutes, which is in agreement with the prediction of chromospheric evaporation by the 1D hydrodynamic flare model. In our study, the velocities of the 
upflow range from 35 to 120 km s$^{-1}$, which is roughly consistent with the results of \citet{pol15}. However, the raster observation of \textit{IRIS} was in the 
``sparse synoptic raster'' mode instead of ``sit and stare'' mode on 2015 October 16, so that we can not study the spectral evolution of the same flare location.

Although we did not carry out magnetic extrapolation based on the photospheric magnetograms since the AR was close to the limb, the CFR surrounding the IFR in Figure~\ref{fig2}(g) and 
the HMI LOS magnetogram in Figure~\ref{fig1}(b) strongly suggest the existence of a null point and spine-fan topology in the corona \citep{mas09,reid12,wang12,jos15}. 
\citet{bau13a} studied the mechanism of particle acceleration in coronal 3D null point reconnection region, finding that sub-relativistic electrons are accelerated by 
a systematic electric field in the current sheet. The impact regions of the high-energy electrons in the chromosphere agree well with previous observations. In this 
work, we are not sure whether the electrons are accelerated by electric field or not. Quantitative calculations are required in the future. Besides, explosive evaporation 
took place at certain locations (see Figure~\ref{fig6}(f)), though the raster covered the whole CFR and IFR. There are credible redshifts in the Si {\sc iv} line at the other locations 
of CFR. However, the intensities of the Fe {\sc xxi} are too weak and the uncertainties of velocities are too large. Magnetic reconnection and particle acceleration 
mechanism are tightly related to the magnetic configuration, and the precipitation of nonthermal electrons along the CFR may not be uniform and isotropic \citep{ros10,bau13a}.
The temperatures of chromosphere at the other locations are probably raised to a few 10$^5$$-$10$^6$ K by limited flux of electrons, which is far less than the 
formation temperatures of Fe {\sc xxi} ($\sim$11 MK) and AIA 94 {\AA} ($>6$ MK) in Figure~\ref{fig3}(h).

\section{Summary} \label{s-sum}
In this paper, we report our multiwavelength observations of the C4.2 circular-ribbon flare by ground-based telescope, \textit{SDO}/AIA, 
\textit{IRIS}, \textit{GOES}, and \textit{RHESSI} on 2015 October 16 in AR 12434. The main results are summarized as follows:
\begin{enumerate}
\item{The short-lived flare was associated with a negative magnetic polarity surrounded by positive polarities in the photosphere, 
which is strongly indicative of a magnetic null point and the fan-spine configuration in the corona.
A mini-filament residing in the AR erupted, generating the IFR and SCFR, which can be considered as a pair of parallel ribbons.}
\item{When the filament reached the null point, it triggered magnetic reconnection with the ambient open field near the null point and generated the closed CFR and a blowout jet. 
The IFR and CFR were cospatial with the negative polarity and positive polarities. The CFR brightening was sequential in the counterclockwise direction in the \textit{IRIS}/SJI images.
The blowout jet moved along the axis at a speed of $\sim$308 km s$^{-1}$. Meanwhile, it drifted from west to east across the axis at a speed of $\sim$87 km s$^{-1}$.}
\item{During the impulsive phase of the flare, there were plasma upflow in the hot Fe {\sc xxi} line at speed of 35$-$120 km s$^{-1}$ 
and downflow in the cool Si {\sc iv} line at speed of 10$-$60 km s$^{-1}$ in the IFR and CFR, indicating explosive chromospheric evaporation during the impulsive phase of flare.}
\item{The IFR was cospatial with the single HXR source at 12$-$25 keV. Calculation based on the thick-target model suggests that the explosive evaporation was 
most probably driven by nonthermal electrons. Whether the electrons were accelerated by electric field in the current sheet during the magnetic reconnection is still unclear.
Additional case studies combined with 3D numerical simulations are required in the future.}
\end{enumerate}

\acknowledgements
The authors appreciate the referee for valuable comments and suggestions. We also acknowledge M. D. Ding, J. Lin, J. X. Cheng, Z. Xu, and H. Tian for inspiring discussions.
\textit{IRIS} is a NASA small explorer mission developed and operated by LMSAL with mission operations executed at NASA Ames Research center 
and major contributions to downlink communications funded by the Norwegian Space Center (NSC, Norway) through an ESA PRODEX contract.
\textit{SDO} is a mission of NASA\rq{}s Living With a Star Program. AIA and HMI data are courtesy of the NASA/\textit{SDO} science teams. 
This work is supported by NSFC No. 11303101, 11333009, 11573072, and the open research 
program of Key Laboratory of Solar Activity, National Astronomical Observatories, CAS No. KLSA201510.
H. S. Ji is supported by the Strategic Priority Research Program$-$The Emergence of Cosmological Structures of the CAS, Grant No. XDB09000000.
Y. N. Su is supported by NSFC 11473071, Youth Fund of Jiangsu BK20141043, and the One Hundred Talent Program of Chinese Academy of Sciences.

\clearpage

\begin{table}
\caption{Description of the observational parameters.}
\label{tbl-1}
\centering
\begin{tabular}{l c c c c}
\hline\hline
Instrument & $\lambda$ & Time & Cadence & Pixel Size \\
           &     ({\AA})   & (UT)  &    (sec)    &  (arcsec)  \\
\hline
  GONG & 6563 & 13:00$-$14:00 & 60  & 1.0 \\
  \textit{SDO}/AIA & 94$-$335 & 13:00$-$14:00 & 12 & 0.6 \\
  \textit{SDO}/AIA & 1600 & 13:00$-$14:00 & 24 & 0.6 \\
  \textit{SDO}/HMI &  6173      & 13:00$-$14:00 & 45 & 0.5 \\
  \textit{IRIS}/SJI & 1400 & 13:09$-$13:56 & $\sim$11.4 & 0.166 \\
  \textit{IRIS}/raster & Fe {\sc xxi} 1354.09 & 13:37:29$-$13:43:00 & $\sim$9.4 & 0.166 \\
  \textit{IRIS}/raster & Si {\sc iv} 1393.77 & 13:37:29$-$13:43:00 & $\sim$9.4 & 0.166 \\
  \textit{GOES}   & 0.5$-$4, 1$-$8  &  13:20$-$14:00 & 2.04  & $-$ \\
  \textit{RHESSI} & 3$-$100 keV & 13:20$-$14:00 & 4, 120  & 4  \\
\hline
\end{tabular}
\end{table}

\clearpage

\clearpage

\clearpage

\clearpage

\clearpage

\clearpage

\clearpage

\clearpage

\clearpage

\clearpage


\begin{thebibliography}{}
\bibitem[Abbett \& Hawley(1999)]{abb99} Abbett, W.~P., \& Hawley, S.~L.\ 1999, \apj, 521, 906
\bibitem[Acton et al.(1982)]{act82} Acton, L.~W., Leibacher, J.~W., Canfield, R.~C., et al.\ 1982, \apj, 263, 409
\bibitem[Allred et al.(2005)]{all05} Allred, J.~C., Hawley, S.~L., Abbett, W.~P., \& Carlsson, M.\ 2005, \apj, 630, 573
\bibitem[Allred et al.(2015)]{all15} Allred, J.~C., Kowalski, A.~F., \& Carlsson, M.\ 2015, \apj, 809, 104
\bibitem[Aschwanden(2004)]{asch04} Aschwanden, M.~J.\ 2004, Physics of the Solar Corona
\bibitem[Battaglia et al.(2009)]{bat09} Battaglia, M., Fletcher, L., \& Benz, A.~O.\ 2009, \aap, 498, 891
\bibitem[Battaglia et al.(2015)]{bat15} Battaglia, M., Kleint, L., Krucker, S., \& Graham, D.\ 2015, \apj, 813, 113
\bibitem[Baumann et al.(2013a)]{bau13a} Baumann, G., Haugb{\o}lle, T., \& Nordlund, {\AA}.\ 2013a, \apj, 771, 93
\bibitem[Baumann et al.(2013b)]{bau13b} Baumann, G., Galsgaard, K., \& Nordlund, {\AA}.\ 2013b, \solphys, 284, 467
\bibitem[Benz(2008)]{benz08} Benz, A.~O.\ 2008, Living Reviews in Solar Physics, 5, 1 
\bibitem[Brannon \& Longcope(2014)]{bra14} Brannon, S., \& Longcope, D.\ 2014, \apj, 792, 50
\bibitem[Brosius \& Phillips(2004)]{bro04} Brosius, J.~W., \& Phillips, K.~J.~H.\ 2004, \apj, 613, 580
\bibitem[Brosius(2009)]{bro09} Brosius, J.~W.\ 2009, \apj, 701, 1209
\bibitem[Brown(1971)]{bro71} Brown, J.~C.\ 1971, \solphys, 18, 489
\bibitem[Canfield et al.(1990)]{can90} Canfield, R.~C., Metcalf, T.~R., Zarro, D.~M., \& Lemen, J.~R.\ 1990, \apj, 348, 333
\bibitem[Czaykowska et al.(1999)]{cza99} Czaykowska, A., De Pontieu, B., Alexander, D., \& Rank, G.\ 1999, \apjl, 521, L75 
\bibitem[Cheng et al.(2010)]{cheng10} Cheng, J.~X., Ding, M.~D., \& Carlsson, M.\ 2010, \apj, 711, 185
\bibitem[De Pontieu et al.(2014)]{dep14} De Pontieu, B., Title, A.~M., Lemen, J.~R., et al.\ 2014, \solphys, 289, 2733
\bibitem[Emslie et al.(1992)]{ems92} Emslie, A.~G., Li, P., \& Mariska, J.~T.\ 1992, \apj, 399, 714
\bibitem[Fisher et al.(1985a)]{fis85a} Fisher, G.~H., Canfield, R.~C., \& McClymont, A.~N.\ 1985, \apj, 289, 414 
\bibitem[Fisher et al.(1985b)]{fis85b} Fisher, G.~H., Canfield, R.~C., \& McClymont, A.~N.\ 1985, \apj, 289, 425
\bibitem[Fisher et al.(1985c)]{fis85c} Fisher, G.~H., Canfield, R.~C., \& McClymont, A.~N.\ 1985, \apj, 289, 434
\bibitem[Fisher(1987)]{fis87} Fisher, G.~H.\ 1987, \apj, 317, 502
\bibitem[Fletcher et al.(2011)]{fle11} Fletcher, L., Dennis, B.~R., Hudson, H.~S., et al.\ 2011, \ssr, 159, 19
\bibitem[Graham \& Cauzzi(2015)]{gra15} Graham, D.~R., \& Cauzzi, G.\ 2015, \apjl, 807, L22
\bibitem[Jiang et al.(2013)]{jia13} Jiang, C., Feng, X., Wu, S.~T., \& Hu, Q.\ 2013, \apjl, 771, L30
\bibitem[Joshi et al.(2015)]{jos15} Joshi, N.~C., Liu, C., Sun, X., et al.\ 2015, \apj, 812, 50
\bibitem[Kumar et al.(2016)]{kum16} Kumar, P., Nakariakov, V.~M., \& Cho, K.-S.\ 2016, \apj, 822, 7
\bibitem[Lau \& Finn(1990)]{lau90} Lau, Y.-T., \& Finn, J.~M.\ 1990, \apj, 350, 672
\bibitem[Lemen et al.(2012)]{lem12} Lemen, J.~R., Title, A.~M., Akin, D.~J., et al.\ 2012, \solphys, 275, 17
\bibitem[Li et al.(2015a)]{li15a} Li, D., Ning, Z.~J., \& Zhang, Q.~M.\ 2015a, \apj, 813, 59
\bibitem[Li et al.(2015b)]{li15b} Li, Y., Ding, M.~D., Qiu, J., \& Cheng, J.~X.\ 2015b, \apj, 811, 7
\bibitem[Li et al.(2016)]{li16} Li, D., Innes, D.~E., \& Ning, Z.~J.\ 2016, \aap, 587, A11
\bibitem[Lin et al.(2002)]{lin02} Lin, R.~P., Dennis, B.~R., Hurford, G.~J., et al.\ 2002, \solphys, 210, 3
\bibitem[Lin et al.(2004)]{lin04} Lin, J., Raymond, J.~C., \& van Ballegooijen, A.~A.\ 2004, \apj, 602, 422
\bibitem[Litvinenko(2004)]{lit04} Litvinenko, Y.~E.\ 2004, The Solar-B Mission and the Forefront of Solar Physics, 325, 355
\bibitem[Liu et al.(2006)]{liu06} Liu, W., Liu, S., Jiang, Y.~W., \& Petrosian, V.\ 2006, \apj, 649, 1124
\bibitem[Liu et al.(2013)]{liu13} Liu, C., Xu, Y., Deng, N., et al.\ 2013, \apj, 774, 60
\bibitem[Liu et al.(2015)]{liu15} Liu, C., Deng, N., Liu, R., et al.\ 2015, \apjl, 812, L19
\bibitem[Longcope(2014)]{long14} Longcope, D.~W.\ 2014, \apj, 795, 10 
\bibitem[Lynch et al.(2009)]{lyn09} Lynch, B.~J., Antiochos, S.~K., Li, Y., Luhmann, J.~G., \& DeVore, C.~R.\ 2009, \apj, 697, 1918
\bibitem[Lynch et al.(2008)]{lyn08} Lynch, B.~J., Antiochos, S.~K., DeVore, C.~R., Luhmann, J.~G., \& Zurbuchen, T.~H.\ 2008, \apj, 683, 1192-1206
\bibitem[Mariska et al.(1989)]{mar89} Mariska, J.~T., Emslie, A.~G., \& Li, P.\ 1989, \apj, 341, 1067
\bibitem[Masson et al.(2009)]{mas09} Masson, S., Pariat, E., Aulanier, G., \& Schrijver, C.~J.\ 2009, \apj, 700, 559
\bibitem[Milligan et al.(2006a)]{mil06a} Milligan, R.~O., Gallagher, P.~T., Mathioudakis, M., \& Keenan, F.~P.\ 2006a, \apjl, 642, L169
\bibitem[Milligan et al.(2006b)]{mil06b} Milligan, R.~O., Gallagher, P.~T., Mathioudakis, M., et al.\ 2006b, \apjl, 638, L117
\bibitem[Moore et al.(2010)]{moo10} Moore, R.~L., Cirtain, J.~W., Sterling, A.~C., \& Falconer, D.~A.\ 2010, \apj, 720, 757
\bibitem[Moore et al.(2013)]{moo13} Moore, R.~L., Sterling, A.~C., Falconer, D.~A., \& Robe, D.\ 2013, \apj, 769, 134
\bibitem[Moreno-Insertis \& Galsgaard(2013)]{more13} Moreno-Insertis, F., \& Galsgaard, K.\ 2013, \apj, 771, 20
\bibitem[Ning et al.(2009)]{ning09} Ning, Z., Cao, W., Huang, J., et al.\ 2009, \apj, 699, 15
\bibitem[Ning \& Cao(2010)]{ning10} Ning, Z., \& Cao, W.\ 2010, \apj, 717, 1232
\bibitem[Parnell et al.(1996)]{par96} Parnell, C.~E., Smith, J.~M., Neukirch, T., \& Priest, E.~R.\ 1996, Physics of Plasmas, 3, 759
\bibitem[Pariat et al.(2009)]{par09} Pariat, E., Antiochos, S.~K., \& DeVore, C.~R.\ 2009, \apj, 691, 61
\bibitem[Polito et al.(2015)]{pol15} Polito, V., Reeves, K.~K., Del Zanna, G., Golub, L., \& Mason, H.~E.\ 2015, \apj, 803, 84
\bibitem[Polito et al.(2016)]{pol16} Polito, V., Reep, J.~W., Reeves, K.~K., et al.\ 2016, \apj, 816, 89
\bibitem[Priest \& Titov(1996)]{pri96} Priest, E.~R., \& Titov, V.~S.\ 1996, Philosophical Transactions of the Royal Society of London Series A, 354, 2951
\bibitem[Priest \& Forbes(2000)]{pri00} Priest, E., \& Forbes, T.\ 2000, Magnetic reconnection : MHD theory and applications / Eric Priest, Terry Forbes.~ New York : Cambridge University Press, 2000
\bibitem[Raftery et al.(2009)]{raf09} Raftery, C.~L., Gallagher, P.~T., Milligan, R.~O., \& Klimchuk, J.~A.\ 2009, \aap, 494, 1127
\bibitem[Reep et al.(2015)]{reep15} Reep, J.~W., Bradshaw, S.~J., \& Alexander, D.\ 2015, \apj, 808, 177
\bibitem[Reep \& Russell(2016)]{reep16} Reep, J.~W., \& Russell, A.~J.~B.\ 2016, \apjl, 818, L20
\bibitem[Reid et al.(2012)]{reid12} Reid, H.~A.~S., Vilmer, N., Aulanier, G., \& Pariat, E.\ 2012, \aap, 547, A52
\bibitem[Rosdahl \& Galsgaard(2010)]{ros10} Rosdahl, K.~J., \& Galsgaard, K.\ 2010, \aap, 511, A73
\bibitem[Sadykov et al.(2015)]{sad15} Sadykov, V.~M., Vargas Dominguez, S., Kosovichev, A.~G., et al.\ 2015, \apj, 805, 167
\bibitem[Scherrer et al.(2012)]{sch12} Scherrer, P.~H., Schou, J., Bush, R.~I., et al.\ 2012, \solphys, 275, 207
\bibitem[Shibata et al.(1992)]{shi92} Shibata, K., Ishido, Y., Acton, L.~W., et al.\ 1992, \pasj, {\bf 44}, L173
\bibitem[Shibata et al.(1995)]{shi95} Shibata, K., Masuda, S., Shimojo, M., et al.\ 1995, \apjl, 451, L83 
\bibitem[Su et al.(2013)]{su13} Su, Y., Veronig, A.~M., Holman, G.~D., et al.\ 2013, Nature Physics, 9, 489 
\bibitem[Sun et al.(2013)]{sun13} Sun, X., Hoeksema, J.~T., Liu, Y., et al.\ 2013, \apj, 778, 139 
\bibitem[Tian et al.(2015)]{tian15} Tian, H., Young, P.~R., Reeves, K.~K., et al.\ 2015, \apj, 811, 139
\bibitem[Wang \& Liu(2012)]{wang12} Wang, H., \& Liu, C.\ 2012, \apj, 760, 101
\bibitem[Wuelser et al.(1994)]{wue94} Wuelser, J.-P., Canfield, R.~C., Acton, L.~W., et al.\ 1994, \apj, 424, 459
\bibitem[Yang et al.(2015)]{yang15} Yang, K., Guo, Y., \& Ding, M.~D.\ 2015, \apj, 806, 171
\bibitem[Yokoyama \& Shibata(1998)]{yoko98} Yokoyama, T., \& Shibata, K.\ 1998, \apjl, 494, L113
\bibitem[Young et al.(2013)]{you13} Young, P.~R., Doschek, G.~A., Warren, H.~P., \& Hara, H.\ 2013, \apj, 766, 127
\bibitem[Zarro \& Lemen(1988)]{zar88} Zarro, D.~M., \& Lemen, J.~R.\ 1988, \apj, 329, 456
\bibitem[Zhang et al.(2012)]{zqm12} Zhang, Q.~M., Chen, P.~F., Guo, Y., Fang, C., \& Ding, M.~D.\ 2012, \apj, 746, 19
\bibitem[Zhang \& Ji(2013)]{zqm13} Zhang, Q.~M., \& Ji, H.~S.\ 2013, \aap, 557, L5
\bibitem[Zhang \& Ji(2014)]{zqm14} Zhang, Q.~M., \& Ji, H.~S.\ 2014, \aap, 561, A134 
\bibitem[Zhang et al.(2015)]{zqm15} Zhang, Q.~M., Ning, Z.~J., Guo, Y., et al.\ 2015, \apj, 805, 4
\end{thebibliography}
\end{document}